\documentclass[aip,reprint]{revtex4-1}%
\usepackage{graphicx}
\usepackage{dcolumn}
\usepackage{bm}
\usepackage{amsmath}%
\usepackage{amsfonts}%
\usepackage{amssymb}
\begin{document}
\bigskip%
\title{Gaussian pulse dynamics in gain media with Kerr nonlinearity}
\author{Christian Jirauschek and Franz X. K\"{a}rtner}
\affiliation{Department of Electrical Engineering and Computer Science
and Research Laboratory of Electronics, Massachusetts Institute of Technology,
77 Massachusetts Avenue, Cambridge, Massachusetts 02139}
\date{14 March 2011, published as J. Opt. Soc. Am. B \textbf{23}, 1776 (2006)}
\begin{abstract}
Using the Kantorovitch method in combination with a Gaussian ansatz, we
derive the equations of motion for spatial, temporal and spatiotemporal
optical propagation in a dispersive Kerr medium with a general transverse
and spectral gain profile. By rewriting the variational equations as
differential equations for the temporal and spatial Gaussian $q$ parameters,
optical $ABCD$ matrices for the Kerr effect, a general transverse gain
profile and nonparabolic spectral gain filtering are obtained. Further
effects can easily be taken into account by adding the corresponding $ABCD$
matrices. Applications include the temporal pulse dynamics in gain fibers
and the beam propagation or spatiotemporal pulse evolution in bulk gain
media.\ As an example, the steady-state spatiotemporal Gaussian pulse
dynamics in a Kerr-lens mode-locked\ laser resonator is studied.
\end{abstract}
\maketitle

\section{INTRODUCTION}

The optical propagation in Kerr media with a transverse and spectral gain
profile is of much interest in many areas. For instance, a combination of gain
guiding and nonlinear self-focusing can play an important role for the spatial
propagation of a laser beam in high-power laser rods. In nonlinear dispersive
gain fibers, e.g., in fiber lasers or optical transmission lines equipped with
erbium-doped fiber amplifiers, the temporal pulse evolution is affected by
self-phase modulation and spectral filtering.\cite{man02} The coupled
spatiotemporal dynamics in the gain medium is relevant for the operation of
Kerr-lens mode-locked (KLM)\ lasers, where the pulse stabilization is governed
by a combination of spatial and temporal effects.\cite{chr98,kal98,jir03}

The variational approach has been extensively used for an approximate
description of the optical propagation in nonlinear media. By describing the
optical field in terms of a trial function with free parameters, a set of
coupled ordinary differential equations can be extracted from the partial
differential equation governing the optical propagation. This allows for an
analytical analysis or an efficient numerical treatment using a standard
differential equation solver. The Rayleigh-Ritz method is a widely-used
variational technique for the treatment of conservative systems, and has been
applied to the spatial, temporal and spatiotemporal optical propagation in
Kerr media.\cite{and79,and79A,and83,and91,jir02} Different approaches have
been developed to include dissipative effects.\cite{kau95,rie96,cer98} Here,
we use a generalization of the Rayleigh-Ritz method known as Kantorovitch
method.\cite{cer98} This technique has for example been applied to the
nonlinear temporal pulse propagation including parabolic spectral gain
filtering,\cite{and99,man02} using a Pereira--Stenflo type ansatz, and to
Gaussian beam propagation in air.\cite{ako00}

In the following, we apply the Kantorovitch method to the description of the
full spatiotemporal optical propagation of Gaussian light bullets in Kerr
media, taking into account an arbitrary gain profile. As temporal effects, we
consider dispersion, self-phase modulation and spectral gain filtering. The
spatial effects include diffraction, self-focusing and a transverse gain
profile. Also the cases of purely spatial beam propagation and purely temporal
pulse evolution are considered. The assumption of a Gaussian ansatz allows us
to relate the equations of motion to the compact and elegant $ABCD$ matrix
formalism for Gaussian beam and pulse propagation.\cite{sie86,dij90} By
rewriting the variational equations as differential equations for the $q$
parameters, we can extract $ABCD$ matrices for the Kerr effect, a general
transverse gain profile and nonparabolic spectral gain filtering. Further
effects, like a parabolic refractive index profile, can easily be incorporated
by adding the corresponding matrix elements.

The paper is organized as follows: In Section \ref{eom}, the spatial, temporal
and spatiotemporal equations of motion for the Gaussian parameters are
obtained from the generalized nonlinear Schr\"{o}dinger equation, which
governs the optical propagation in Kerr media. In Section \ref{ABCD}, these
equations are reformulated within the framework of the $ABCD$ matrix
formalism, taking advantage of the Gaussian $q$ parameter description. As an
example, the Gaussian pulse dynamics in a KLM\ laser resonator is studied in
Section \ref{example}, including a soft gain aperture and spectral filtering.
The paper is concluded in Section \ref{conclusion}.

\section{\label{eom}VARIATIONAL APPROACH}

A linearly polarized light pulse, propagating in $z$ direction through a
dispersive Kerr medium with a parabolic transverse and spectral gain profile,
is described by the generalized nonlinear Schr\"{o}dinger equation%
\begin{equation}
\mathrm{i}\partial_{z}U-\mathcal{D}\partial_{t_{\mathrm{r}}}^{2}%
U+\mathcal{B}\left(  \partial_{x}^{2}+\partial_{y}^{2}\right)  U+\delta\left|
U\right|  ^{2}U=Q \label{nls}%
\end{equation}
with the gain term%
\begin{equation}
Q=\mathrm{i}\left(  g_{0}-g_{x}x^{2}-g_{y}y^{2}+g_{\omega}\partial
_{t_{\mathrm{r}}}^{2}\right)  U. \label{Q}%
\end{equation}
The retarded time is defined as $t_{\mathrm{r}}=t-z/v_{\mathrm{g}}\ $with the
group velocity $v_{\mathrm{g}}$, $x$ and $y$ are the transverse coordinates.
$U$ is the slowly varying envelope, normalized such that its absolute square
gives the intensity of the wave. The transverse electrical field component
$E_{\mathrm{t}}$ is related to $U$ by $E_{\mathrm{t}}=\sqrt{2Z_{0}/n_{0}%
}\allowbreak\times\Re\{U\exp\left[  \mathrm{i}\left(  kz-\omega_{0}t\right)
\right]  \}$, where $n_{0}$ and $k=n_{0}k_{0}$ are the refractive index and
the wavenumber at the center frequency $\omega_{0}$, and $Z_{0}$ is the wave
resistance in vacuum. Here we use the `physics' convention, in which a plane
wave is described by $\exp\left[  \mathrm{i}\left(  kz-\omega_{0}t\right)
\right]  $. The `engineering' notation $\exp\left[  \mathrm{j}\left(
\omega_{0}t-kz\right)  \right]  $ can easily be obtained by the formal
transcription $\mathrm{i}\rightarrow-\mathrm{j}$ in all expressions. The
parameters for dispersion and diffraction are given by $\mathcal{D}=\frac
{1}{2}k^{\prime\prime}$, where $k^{\prime\prime}$ is the second derivative of
the wavenumber at $\omega_{0}$, and $\mathcal{B}=1/\left(  2k\right)  $. The
nonlinearity parameter is $\delta=k_{0}n_{2I}$, where $n_{2I}$ is the
intensity dependent refractive index, so that the total refractive index is
given by $n=n_{0}+n_{2I}\left|  U\right|  ^{2}$. In general, the coefficients
$\mathcal{D}$, $\mathcal{B}$, $\delta$, $g_{0}$, $g_{x}$, $g_{y}$, $g_{\omega
}$, and thus $n_{0}$ and $k^{\prime\prime}$, depend on the position $z$ in the
medium. For example, the material parameters change abruptly at the interface
between two materials, and in optically pumped gain media, $g_{0}$, $g_{x}$
and $g_{y}$ depend on $z$ due to the divergence of the pump beam. Although the
$z$ argument is suppressed for a more compact notation, all the equations
given in this paper are valid for $z$ dependent coefficients.

As a trial function for $U$, we choose a complex Gaussian, which has been
widely used for the variational analysis of optical propagation in Kerr
media.\cite{and79,and83,and91,jir02} Since it is an exact solution of
Eq.\thinspace(\ref{nls}) for $\delta=0$, we expect it to be a good
approximation to the exact solution, at least for moderate nonlinearity. It
has been shown that for the temporal dispersion-managed soliton dynamics, the
Gaussian description is applicable over a wide parameter
range.\cite{chen99,chen99A} The transverse Gaussian field distribution, which
is the fundamental mode in linear paraxial resonators, is widely used as an
approximate description for the transverse field distribution in nonlinear
resonators.\cite{pen96,kal98,jir03} In addition, the Gaussian trial function
can conveniently be characterized in terms of the complex $q$ parameters,
which allow for a compact description of the optical propagation based on the
$ABCD$ matrix formalism.\cite{sie86,dij90}

In the following, the Gaussian equations of motion, obtained by the
Kantorovitch method, are given for the spatial beam propagation and the
temporal pulse evolution in a Kerr medium with a parabolic transverse and
spectral gain profile, as well as for the full spatiotemporal dynamics. The
derivation of the spatiotemporal equations can be found in Appendix A; the
purely spatial and temporal equations are obtained in an analogous manner.

\subsection{Equations of Motion}

The generalized nonlinear Schr\"{o}dinger equation, as given in Eq.\thinspace
(\ref{nls}), describes the spatiotemporal pulse dynamics in a dispersive Kerr
medium, for example the propagation of a light bullet in a Kerr-lens
mode-locked (KLM)\ laser, taking into account a parabolic transverse and
spectral gain profile. The Gaussian test function is given by%
\begin{align}
U(z,t_{\mathrm{r}},x,y) &  =\hat{U}(z)\exp\bigg\{-\left[  \frac{1}{2T^{2}%
(z)}-\mathrm{i}b(z)\right]  t_{\mathrm{r}}^{2}\nonumber\\
&  -\left[  \frac{1}{2w_{x}^{2}(z)}-\mathrm{i}a_{x}(z)\right]  x^{2}%
\nonumber\\
&  -\left[  \frac{1}{2w_{y}^{2}(z)}-\mathrm{i}a_{y}(z)\right]  y^{2}\bigg
\},\label{Uwt_wt}%
\end{align}
with the beam widths $w_{x}$ and $w_{y}$, the pulse duration $T$, the spatial
chirp parameters $a_{x}$ and $a_{y}$, and the temporal chirp parameter $b$.
The complex amplitude can be written as%
\begin{equation}
\hat{U}(z)=A(z)\exp\left[  \mathrm{i}\phi(z)\right]  .\label{U}%
\end{equation}
The derivation of the equations of motion is given in Appendix A. The
equations for the beam width and the pulse duration are given by
\begin{subequations}
\label{wT}%
\begin{align}
w_{p}^{\prime} &  =4\mathcal{B}a_{p}w_{p}-g_{p}w_{p}^{3}{},\label{w}\\
T^{\prime} &  =-4\mathcal{D}bT+g_{\omega}\left(  \frac{1}{T}-4b^{2}%
T^{3}\right)  ,\label{T}%
\end{align}
where $p=x,y$, and the prime denotes a partial derivative with respect to $z$.
Taking the full spatiotemporal dynamics into account, the intensity dependent
contributions in the equations for the chirp parameters and the phase have
different prefactors $c_{a,\phi}$, as compared to the purely temporal or
spatial dynamics. For the chirp parameters, we obtain
\end{subequations}
\begin{subequations}
\label{ab}%
\begin{align}
a_{p}^{\prime} &  =\mathcal{B}\left(  \frac{1}{w_{p}^{4}}-4a_{p}^{2}\right)
-c_{a}\delta\frac{A^{2}}{w_{p}^{2}},\label{a}\\
b^{\prime} &  =-\mathcal{D}\left(  \frac{1}{T^{4}}-4b^{2}\right)  -4g_{\omega
}\frac{b}{T^{2}}-c_{a}\delta\frac{A^{2}}{T^{2}},\label{b}%
\end{align}
with $p=x,y$ and $c_{a}=\sqrt{2}/8$. For the amplitude, the variational
principle yields
\end{subequations}
\begin{equation}
A^{\prime}=A\left(  g_{0}-\frac{g_{\omega}}{T^{2}}+2\mathcal{D}b-2\mathcal{B}%
a_{x}-2\mathcal{B}a_{y}\right)  ,\label{A_wt}%
\end{equation}
and the phase evolution is described by%
\begin{equation}
\phi^{\prime}=2g_{\omega}b+\mathcal{D}\frac{1}{T^{2}}-\mathcal{B}\frac
{1}{w_{x}^{2}}-\mathcal{B}\frac{1}{w_{y}^{2}}+c_{\phi}\delta A^{2}%
\label{phi_wt}%
\end{equation}
with $c_{\phi}=7\sqrt{2}/16$.

Setting $\mathcal{D}=g_{\omega}=0$ in Eqs.\thinspace(\ref{nls}) and (\ref{Q})
yields the nonlinear Sch\"{o}dinger equation for the purely spatial dynamics.
This equation describes the cw propagation of a beam with a transverse field
distribution $U=U\left(  z,x,y\right)  $ in a Kerr medium with a transverse
gain profile, for example the nonlinear gain medium of an optically pumped
solid-state laser. The Gaussian trial function is given by
\begin{align}
U(z,x,y) &  =\hat{U}(z)\exp\bigg\{-\left[  \frac{1}{2w_{x}^{2}(z)}%
-\mathrm{i}a_{x}(z)\right]  x^{2}\nonumber\\
&  -\left[  \frac{1}{2w_{y}^{2}(z)}-\mathrm{i}a_{y}(z)\right]  y^{2}\bigg
\},\label{Uw_w}%
\end{align}
with the complex amplitude defined in Eq.\thinspace(\ref{U}). The relevant
equations of motion are here Eqs.\thinspace(\ref{w}), (\ref{a}), (\ref{A_wt})
and (\ref{phi_wt}) with $\mathcal{D}=g_{\omega}=0$. The nonlinearity
coefficients are now given by $c_{a}=1/4$ and $c_{\phi}=3/4$, which are the
same as derived using the method of minimum weighted square mean
error,\cite{mag93A} but different from the results obtained by a Taylor expansion.\cite{pen96}%

\begin{table*}[t]
\caption{Gaussian test function, relevant equations of motion and coefficients
for spatiotemporal, spatial and temporal dynamics.}
\begin{center}
\begin{tabular}
[c]{llll}\hline
Propagation & Pulse & Equations of Motion & Coefficients\\\hline
Spatiotemporal & Eq. (\ref{Uwt_wt}) & Eqs. (\ref{wT}),(\ref{ab}),(\ref{A_wt}
),(\ref{phi_wt}) & $c_{a}=\sqrt{2}/8,c_{\phi}=7\sqrt{2}/16$\\
Spatial & Eq. (\ref{Uw_w}) & Eqs. (\ref{w}),(\ref{a}),(\ref{A_wt}
),(\ref{phi_wt}) & $c_{a}=1/4,c_{\phi}=3/4,\mathcal{D}=g_{\omega}=0$\\
Temporal & Eq. (\ref{Ut_t}) & Eqs. (\ref{T}),(\ref{b}),(\ref{A_wt}
),(\ref{phi_wt}) & $c_{a}=\sqrt{2}/4,c_{\phi}=5\sqrt{2}/8,\mathcal{B}
=g_{x,y}=0$\\\hline\end{tabular}
\end{center}
\end{table*}%

The propagation equations for the purely temporal pulse dynamics are obtained
by setting $\mathcal{B}=g_{x}=g_{y}=0$ in Eqs.\thinspace(\ref{nls}) and
(\ref{Q}). This equation describes the propagation of a pulse $U=U\left(
z,t_{\mathrm{r}}\right)  $ in a nonlinear dispersive medium with frequency
dependent loss or gain, like an optical amplifier. This equation is also
referred to as complex cubic Ginzburg--Landau equation. The temporal Gaussian
pulse shape is described by%
\begin{equation}
U(z,t_{\mathrm{r}})=\hat{U}(z)\exp\left\{  -\left[  \frac{1}{2T^{2}%
(z)}-\mathrm{i}b(z)\right]  t_{\mathrm{r}}^{2}\right\}  ,\label{Ut_t}%
\end{equation}
with the complex amplitude $\hat{U}$, see Eq.\thinspace(\ref{U}). The relevant
equations of motion are here given by Eqs.\thinspace(\ref{T}), (\ref{b}),
(\ref{A_wt}) and (\ref{phi_wt}) with $\mathcal{B}=0$ and the nonlinearity
coefficients $c_{a}=\sqrt{2}/4$, $c_{\phi}=5\sqrt{2}/8$, which are the same as
obtained by the method of minimum weighted square mean error.\cite{lar00}
Table 1 contains an overview of the suitable Gaussian test function, the
relevant equations of motion and the coefficients for the spatiotemporal,
spatial and temporal dynamics.

For $\delta=0$, the Gaussian is an exact solution, and thus the variational
principle yields the exact result. For $\delta\neq0$, the Kerr nonlinearity
results in an additional intensity dependent spatial and temporal chirp, see
Eq.\thinspace(\ref{ab}), and phase shift, see Eq.\thinspace(\ref{phi_wt}).
Eqs.\thinspace(\ref{wT})--(\ref{phi_wt}) look rather complicated. However, the
physics can be made rather obvious by casting those formulas into mapping
matrices for the complex $q$ parameters of Gaussian bullets, see Section
\ref{ABCD}. In the following, we extend the equations of motion to take into
account arbitrary gain profiles by introducing effective parabolic gain
coefficients. In this case, the Gaussian is not an exact solution of
Eq.\thinspace(\ref{nls}) even if $\delta=0$.

\subsection{\label{np}Nonparabolic Gain Profile}

The equations of motion for the parabolic gain profile given in Eq.\thinspace
(\ref{Q}) can be modified to describe a general transverse and spectral gain
dependence $g\left(  z,\omega,x,y\right)  $, where $\omega$ is a relative
frequency coordinate, centered around the carrier frequency $\omega_{0}$. The
inherent symmetry properties of the Gaussian ansatz make it particularly
suited for describing the propagation in media with gain profiles which are
symmetric around $x=0$, $y=0$ and $\omega=0$. Then a second order Taylor
expansion of $g$ results in a parabolic gain profile. However, this parabolic
approximation is only viable if the transverse and spectral pulse width is
narrow as compared to the gain profile. In laser media, where a large spatial
overlap with the gain is desired, this assumption generally fails. Also the
spectral pulse width can significantly exceed the gain bandwidth, especially
for few-cycle laser pulses.

The gain term of the nonlinear Schr\"{o}dinger equation, Eq.\thinspace
(\ref{nls}), is here given by%
\begin{equation}
Q=\mathrm{i}\mathcal{F}_{t}^{-1}\left\{  g\mathcal{F}_{t}\left\{  U\right\}
\right\}  ,\label{Q_g}%
\end{equation}
with the definition of the Fourier transform%
\begin{equation}
\mathcal{F}_{t}\left\{  U\right\}  =\tilde{U}=\int_{-\infty}^{\infty
}\mathrm{d}t_{\mathrm{r}}\,U\exp\left(  \mathrm{i}\omega t_{\mathrm{r}%
}\right)  .\label{four}%
\end{equation}
For general gain profiles, the Gaussian test function is only an approximate
solution even for $\delta=0$. In Appendix A, the Kantorovitch method is used
to extract the equations of motion for a general gain profile, which can be
brought into the form Eqs.\thinspace(\ref{wT})\-\---(\ref{phi_wt}) by defining
$g_{\omega,x,y}$ and $g_{0}$ as functions of the position dependent Gaussian
parameters:
\begin{subequations}
\label{gwt}%
\begin{align}
g_{\omega} &  =\frac{1}{2\pi}\frac{1}{\Omega^{4}E}\int_{-\infty}^{\infty}%
\int_{-\infty}^{\infty}\int_{-\infty}^{\infty}\left(  \Omega^{2}-2\omega
^{2}\right)  g\left|  \tilde{U}\right|  ^{2}\mathrm{d}\omega\mathrm{d}%
x\mathrm{d}y,\\
g_{x} &  =\frac{1}{2\pi}\frac{1}{w_{x}^{4}E}\int_{-\infty}^{\infty}%
\int_{-\infty}^{\infty}\int_{-\infty}^{\infty}\left(  w_{x}^{2}-2x^{2}\right)
g\left|  \tilde{U}\right|  ^{2}\mathrm{d}\omega\mathrm{d}x\mathrm{d}y,\\
g_{y} &  =\frac{1}{2\pi}\frac{1}{w_{y}^{4}E}\int_{-\infty}^{\infty}%
\int_{-\infty}^{\infty}\int_{-\infty}^{\infty}\left(  w_{y}^{2}-2y^{2}\right)
g\left|  \tilde{U}\right|  ^{2}\mathrm{d}\omega\mathrm{d}x\mathrm{d}y,\\
g_{0} &  =\frac{1}{2\pi}\frac{1}{E}\int_{-\infty}^{\infty}\int_{-\infty
}^{\infty}\int_{-\infty}^{\infty}g\left|  \tilde{U}\right|  ^{2}%
\mathrm{d}\omega\mathrm{d}x\mathrm{d}y\nonumber\\
&  +\frac{g_{x}}{2}w_{x}^{2}+\frac{g_{y}}{2}w_{y}^{2}+\frac{g_{\omega}}%
{2}\Omega^{2},
\end{align}
with the pulse energy $E=\pi^{3/2}A^{2}Tw_{x}w_{y}$ and the spectral $1/e$
width defined as
\end{subequations}
\begin{equation}
\Omega=\sqrt{\frac{1}{T^{2}}+4b^{2}T^{2}}.\label{Om}%
\end{equation}
Here, $\left|  \tilde{U}\right|  ^{2}$ is given by%
\begin{equation}
\left|  \tilde{U}\right|  ^{2}=A^{2}\frac{2\pi T}{\Omega}\exp\left(
-x^{2}/w_{x}^{2}-y^{2}/w_{y}^{2}-\omega^{2}/\Omega^{2}\right)  .\label{F}%
\end{equation}
Eq.\thinspace(\ref{gwt}) provides a position dependent effective parabolic
profile, which depends on $g$ as well as the spectral and transverse pulse
widths. If we are interested in the purely spatial propagation of a Gaussian
beam, Eq.\thinspace(\ref{Uw_w}), in a medium with a gain profile $g\left(
z,x,y\right)  $, we have to use the effective parabolic gain parameters
\begin{subequations}
\label{gw}%
\begin{align}
g_{x} &  =\frac{1}{w_{x}^{4}{}P}\int_{-\infty}^{\infty}\int_{-\infty}^{\infty
}\left(  w_{x}^{2}-2x^{2}\right)  g\left|  U\right|  ^{2}\mathrm{d}%
x\mathrm{d}y,\\
g_{y} &  =\frac{1}{w_{y}^{4}{}P}\int_{-\infty}^{\infty}\int_{-\infty}^{\infty
}\left(  w_{y}^{2}-2y^{2}\right)  g\left|  U\right|  ^{2}\mathrm{d}%
x\mathrm{d}y,\\
g_{0} &  =\frac{1}{P}\int_{-\infty}^{\infty}\int_{-\infty}^{\infty}g\left|
U\right|  ^{2}\mathrm{d}x\mathrm{d}y+\frac{g_{x}}{2}w_{x}^{2}+\frac{g_{y}}%
{2}w_{y}^{2},
\end{align}
with the power $P=\pi A^{2}w_{x}w_{y}$ and
\end{subequations}
\begin{equation}
\left|  U\right|  ^{2}=A^{2}\exp\left(  -x^{2}/w_{x}^{2}-y^{2}/w_{y}%
^{2}\right)  .
\end{equation}
For the purely temporal dynamics of a Gaussian pulse, Eq.\thinspace
(\ref{Ut_t}), the effective parabolic gain parameters for a gain profile
$g\left(  z,\omega\right)  $ are given by
\begin{subequations}
\label{gt}%
\begin{align}
g_{\omega} &  =\frac{1}{2\pi}\frac{1}{\Omega^{4}F}\int_{-\infty}^{\infty
}\left(  \Omega^{2}-2\omega^{2}\right)  g\left|  \tilde{U}\right|
^{2}\mathrm{d}\omega,\\
g_{0} &  =\frac{1}{2\pi}\frac{1}{F}\int_{-\infty}^{\infty}g\left|
\tilde{U}\right|  ^{2}\mathrm{d}\omega+\frac{g_{\omega}}{2}\Omega^{2},
\end{align}
with the fluence $F=\pi^{1/2}A^{2}T$ and
\end{subequations}
\begin{equation}
\left|  \tilde{U}\right|  ^{2}=A^{2}\frac{2\pi T}{\Omega}\exp\left(
-\omega^{2}/\Omega^{2}\right)  .
\end{equation}

As an example, let's consider a Gaussian gain profile%
\begin{equation}
g=\hat{g}\exp\left(  -\Delta_{x}^{-2}x^{2}-\Delta_{y}^{-2}y^{2}-\Delta
_{\omega}^{-2}\omega^{2}\right)  ,\label{gg}%
\end{equation}
with the transverse $1/e$ gain widths $\Delta_{x}$ and $\Delta_{y}$ and the
$1/e$ gain bandwidth $\Delta_{\omega}$. The equations for the effective
parabolic gain profile, Eq.\thinspace(\ref{gwt}), evaluate to
\begin{align}
g_{\omega} &  =\hat{g}\Delta_{\omega}^{-2}\left(  \Delta_{\omega}^{-2}%
\Omega^{2}+1\right)  ^{-3/2}\left(  \Delta_{x}^{-2}w_{x}^{2}+1\right)
^{-1/2}\nonumber\\
&  \times\left(  \Delta_{y}^{-2}w_{y}^{2}+1\right)  ^{-1/2},\nonumber\\
g_{x} &  =\hat{g}\Delta_{x}^{-2}\left(  \Delta_{\omega}^{-2}\Omega
^{2}+1\right)  ^{-1/2}\left(  \Delta_{x}^{-2}w_{x}^{2}+1\right)
^{-3/2}\nonumber\\
&  \times\left(  \Delta_{y}^{-2}w_{y}^{2}+1\right)  ^{-1/2},\nonumber\\
g_{y} &  =\hat{g}\Delta_{y}^{-2}\left(  \Delta_{\omega}^{-2}\Omega
^{2}+1\right)  ^{-1/2}\left(  \Delta_{x}^{-2}w_{x}^{2}+1\right)
^{-1/2}\label{g_wt}\\
&  \times\left(  \Delta_{y}^{-2}w_{y}^{2}+1\right)  ^{-3/2},\nonumber\\
g_{0} &  =\frac{1}{6}\big[g_{\omega}\left(  5\Omega^{2}+2\Delta_{\omega}%
^{2}\right)  +g_{x}\left(  5w_{x}^{2}+2\Delta_{x}^{2}\right)  \nonumber\\
&  +g_{y}\left(  5w_{y}^{2}+2\Delta_{y}^{2}\right)  \big].\nonumber
\end{align}
With the transverse gain profile%
\begin{equation}
g=\hat{g}\exp\left(  -\Delta_{x}^{-2}x^{2}-\Delta_{y}^{-2}y^{2}\right)  ,
\end{equation}
Eq.\thinspace(\ref{gw}) for the spatial beam propagation gives%
\begin{align}
g_{x} &  =\hat{g}\Delta_{x}^{-2}\left(  \Delta_{x}^{-2}w_{x}^{2}+1\right)
^{-3/2}\left(  \Delta_{y}^{-2}w_{y}^{2}+1\right)  ^{-1/2},\nonumber\\
g_{y} &  =\hat{g}\Delta_{y}^{-2}\left(  \Delta_{x}^{-2}w_{x}^{2}+1\right)
^{-1/2}\left(  \Delta_{y}^{-2}w_{y}^{2}+1\right)  ^{-3/2},\label{g_w}\\
g_{0} &  =g_{x}\left(  w_{x}^{2}+\Delta_{x}^{2}/2\right)  +g_{y}\left(
w_{y}^{2}+\Delta_{y}^{2}/2\right)  ,\nonumber
\end{align}
and with the spectral gain profile%
\begin{equation}
g=\hat{g}\exp\left(  -\Delta_{\omega}^{-2}\omega^{2}\right)  ,
\end{equation}
Eq.\thinspace(\ref{gt}) for the temporal pulse propagation becomes%
\begin{align}
g_{\omega} &  =\hat{g}\Delta_{\omega}^{-2}\left(  \Delta_{\omega}^{-2}%
\Omega^{2}+1\right)  ^{-3/2},\nonumber\\
g_{0}\allowbreak &  =g_{\omega}\left(  \frac{3}{2}\Omega^{2}+\Delta_{\omega
}^{2}\right)  .\label{g_t}%
\end{align}

The implementation of a general gain profile opens up the possibility to
include a broad range of saturation effects. For example, the equations of
motion can be coupled to a differential equation for the gain profile $g$,
describing the evolution of $g$ in dependence on the pulse parameters.

\section{\label{ABCD}DESCRIPTION BY OPTICAL MATRICES}

It is convenient to recast the equations of motion in a form consistent with
the familiar $q$ parameter analysis for Gaussian optical propagation, where
the optical elements are described by matrices.\cite{sie86,dij90,kos90} This
notation is compact and very practical, since it allows for a straightforward
treatment of the successive propagation through linear optical elements and
nonlinear Kerr media. Additional effects in the Kerr medium, like a parabolic
refractive index profile, can be easily incorporated by adding the
corresponding matrix elements. On the other hand, by rewriting the variational
equations as differential equations for the $q$ parameters, $ABCD$ matrices
for the Kerr effect and a nonparabolic gain profile can be extracted. We note
that the matrices for the spatiotemporal Kerr effect obtained here are
different from the ones derived previously based on a Taylor expansion
approach, which significantly overestimates the nonlinearity.\cite{chi93}

Using the complex $q$ parameters $q_{x}$, $q_{y}$ and $q_{t}$, we can write
the spatiotemporal Gaussian ansatz as%
\begin{equation}
U(z,t_{\mathrm{r}},x,y)=\hat{U}(z)\exp\left[  \frac{\mathrm{i}k_{0}x^{2}%
}{2q_{x}(z)}+\frac{\mathrm{i}k_{0}y^{2}}{2q_{y}(z)}-\frac{\mathrm{i}\omega
_{0}t_{\mathrm{r}}^{2}}{2q_{t}(z)}\right]  .\label{Uq_wt}%
\end{equation}
Comparison with Eq.\thinspace(\ref{Uwt_wt}) yields%
\begin{equation}
q_{p}^{-1}=\frac{1}{k_{0}}\left(  \frac{\mathrm{i}}{w_{p}^{2}}+2a_{p}\right)
\label{qp}%
\end{equation}
with $p=x,y$, and%
\begin{equation}
q_{t}^{-1}=-\frac{1}{\omega_{0}}\left(  \frac{\mathrm{i}}{T^{2}}+2b\right)
.\label{qt}%
\end{equation}
Here, $q_{x}$ and $q_{y}$ are the reduced $q$ parameters,\cite{sie86} with the
vacuum wavenumber $k_{0}$ in the exponent of Eq.\thinspace(\ref{Uq_wt}),
instead of the wavenumber in the medium. This definition has the advantage
that a $z$ dependent refractive index does not lead to additional terms in the
equations of motion for the $q$ parameter formalism. Likewise, $q_{t}$ is the
temporal analogon to the spatial reduced $q$ parameters,\cite{dij90} with
$\omega_{0}$ in the exponent, instead of the dispersion coefficient. Here, the
dispersion is allowed to depend on the position $z$, without modifications of
the equations.

With the $q$ parameters, the Gaussian beam profile for purely spatial
propagation, Eq.\thinspace(\ref{Uw_w}), can be written as%
\begin{equation}
U=\hat{U}\exp\left(  \frac{\mathrm{i}k_{0}x^{2}}{2q_{x}}+\frac{\mathrm{i}%
k_{0}y^{2}}{2q_{y}}\right)  ,\label{Uq_w}%
\end{equation}
and the temporal Gaussian pulse for describing the purely teporal dynamics,
Eq.\thinspace(\ref{Ut_t}), becomes
\begin{equation}
U=\hat{U}\exp\left(  -\frac{\mathrm{i}\omega_{0}t_{\mathrm{r}}^{2}}{2q_{t}%
}\right)  .\label{Uq_t}%
\end{equation}
In Appendix B it is shown that the propagation equations for the $q$
parameters and the amplitude can be written as coupled differential equations%
\begin{equation}
\partial_{z}q_{s}=-q_{s}^{2}C_{s}^{\prime}+B_{s}^{\prime}\label{q2_w}%
\end{equation}
with $s=x,y,t$, and%
\begin{equation}
\partial_{z}\hat{U}=\hat{U}\left(  -\frac{B_{t}^{\prime}}{2q_{t}}-\frac
{B_{x}^{\prime}}{2q_{x}}-\frac{B_{y}^{\prime}}{2q_{y}}+\alpha\right)
.\label{U2_wt}%
\end{equation}
\ The coefficients $B_{s}^{\prime}$ and $C_{s}^{\prime}$, which in general
depend on the position $z$, can be interpreted as elements in an $ABCD$ matrix
of the form%
\begin{equation}
M_{s}=\left(
\begin{array}
[c]{cc}%
1 & B_{s}^{\prime}\delta z\\
C_{s}^{\prime}\delta z & 1
\end{array}
\right)  ,\label{Ms}%
\end{equation}
describing the propagation in the medium through an infinitely small section
with length $\delta z$. In a gain medium with Kerr nonlinearity, they are
given by%

\begin{align}
B_{p}^{\prime} &  =2\mathcal{B}k_{0}=\frac{1}{n_{0}},\nonumber\\
B_{t}^{\prime} &  =2\mathcal{D}\omega_{0}+2\mathrm{i}\omega_{0}g_{\omega
},\nonumber\\
C_{p}^{\prime} &  =\frac{2\mathrm{i}g_{p}}{k_{0}}-\frac{2c_{a}\delta}%
{k_{0}w_{p}^{2}}\left|  \hat{U}\right|  ^{2},\label{BC}\\
C_{t}^{\prime} &  =\frac{2c_{a}\delta}{\omega_{0}T^{2}}\left|  \hat{U}\right|
^{2},\nonumber
\end{align}
with $p=x,y$, $w_{p}=\left(  k_{0}\Im\left\{  q_{p}^{-1}\right\}  \right)
^{-1/2}$, $T=\left(  -\omega_{0}\Im\left\{  q_{t}^{-1}\right\}  \right)
^{-1/2}$, and the nonlinearity coefficients listed in Table 1. For the complex
on-axis transmission coefficient, we obtain
\begin{equation}
\alpha=g_{0}+\mathrm{i}c_{\phi}\allowbreak\delta\left|  \hat{U}\right|
^{2}.\label{alpha}%
\end{equation}
The purely spatial propagation equations for a Gaussian beam, Eq.\thinspace
(\ref{Uq_w}), are given by Eq.\thinspace(\ref{q2_w}) with $s=x,y$ and
Eq.\thinspace(\ref{U2_wt}) with $B_{t}^{\prime}=0$. The evolution of a purely
temporal Gaussian pulse, Eq.\thinspace(\ref{Uq_t}), is described by
Eq.\thinspace(\ref{q2_w}) with $s=t$ and Eq.\thinspace(\ref{U2_wt}) with
$B_{x}^{\prime}=B_{y}^{\prime}=0$.

\begin{table}[h]
\caption{Optical matrix elements for spatiotemporal Gaussian pulse propagation
through Kerr media with a spatial and spectral gain profile. $B_{t}^{\prime
}=C_{t}^{\prime}=0$ for purely spatial beam propagation; likewise,
$B_{p}^{\prime}=C_{p}^{\prime}=0$ for purely temporal pulse propagation. The
Kerr parameters $c_{a}$ and $c_{\phi}$ are listed in Table 1. A general
nonparabolic gain profile can be taken into account by using the $g_{\omega}$,
$g_{p}$ and $g_{0}$ given in Eqs. (\ref{gwt}), (\ref{gw}) and (\ref{gt}) for
spatiotemporal, purely spatial and purely temporal propagation, respectively.
}
\begin{center}%
\begin{tabular}
[c]{llllll}\hline
Effect & $B_{p}^{\prime}$ & $C_{p}^{\prime}$ & $B_{t}^{\prime}$ &
$C_{t}^{\prime}$ & $\alpha$\\\hline
Free space & $n_{0}^{-1}$ & $0$ &  &  & $0$\\
Dispersion &  &  & $2\mathcal{D}\omega_{0}$ & $0$ & $0$\\
Parabolic gain & $0$ & $2\mathrm{i}g_{p}k_{0}^{-1}$ & $2\mathrm{i}g_{\omega
}\omega_{0}$ & $0$ & $g_{0}$\\
Kerr effect & $0$ & $-c_{a}\frac{2\delta}{k_{0}w_{p}^{2}}|\hat{U}|^{2}$ & $0$
& $c_{a}\frac{2\delta}{\omega_{0}T^{2}}|\hat{U}|^{2}$ & $\mathrm{i}c_{\phi
}\delta|\hat{U}\!|^{2}$\\\hline
\end{tabular}
\end{center}
\end{table}

The coefficients are composed of various contributions, $B_{s}^{\prime}%
=\sum_{i}B_{s,i}^{\prime}$, $C_{s}^{\prime}=\sum_{i}C_{s,i}^{\prime}$, where
the index $i$ represents the different physical mechanisms affecting the
propagation. Each of these effects itself can be described by a matrix of the
form Eq.\thinspace(\ref{Ms}). We identify the matrix elements for free space
propagation ($B_{p,1}=n_{0}^{-1}\delta z$ and $C_{p,1}=0$), soft aperturing
($B_{p,2}=0$ and $C_{p,2}=2\mathrm{i}g_{p}k_{0}^{-1}\delta z$), dispersion
($B_{t,1}=2\mathcal{D}\omega_{0}\delta z$ and $C_{t,1}=0$), and spectral
parabolic filtering ($B_{t,2}=2\mathrm{i}\omega_{0}g_{\omega}\delta z$ and
$C_{t,2}=0$).\cite{sie86,dij90,nak98} The Kerr effect is incorporated as an
intensity dependent lens \cite{mag93A} in the spatial domain, describing the
self-focusing action, and an intensity dependent ''chirper'' or temporal lens
\cite{kol89} in the time domain for the self-phase modulation. An overview of
the matrix elements for the different effects is given in Table 2.\ Note that
a general gain profile deviating from the parabolic approximation can be
incorporated by using above gain matrix elements together with the equations
for $g_{p}$, $g_{\omega}$ and $g_{0}$ given in Section \ref{np}.

Additional effects can easily be incorporated by adding further matrix
elemens. For instance, a parabolic refractive index profile $n\left(
r\right)  =n_{0}-\left(  n_{2,x}x^{2}+n_{2,y}y^{2}\right)  /2$, as generated
by thermal lensing in a laser rod, can be taken into account by the elements
$B_{p}^{\prime}=0$, $C_{p}^{\prime}=-n_{2,p}$. Also transversely varying
saturable gain can be treated by additional $ABCD$ matrices.\cite{gra01}

\section{\label{example}EXAMPLE}

An important application of the Gaussian approximation described above is the
simulation of the optical propagation in laser resonators. Examples are the
temporal pulse evolution in mode-locked fiber lasers, the laser beam
propagation in high-power laser rods, or the spatiotemporal pulse\ dynamics in
Kerr-lens mode-locked lasers. Here, the steady-state solutions cannot be
obtained directly from the optical matrices, since the matrix elements for the
Kerr nonlinearity and the general gain profile depend on the pulse parameters.
Rather, the steady state must be found by iteratively solving the equations,
i.e., propagation over many roundtrips till the steady state is reached.

\begin{figure}[th]
\centering\includegraphics{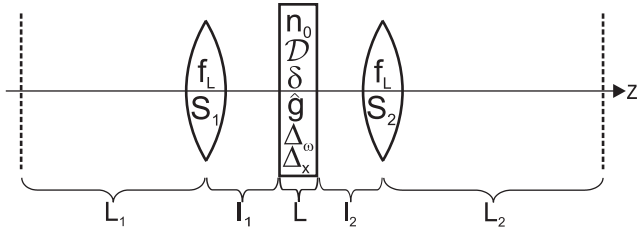}\caption{Simplified model of a
Kerr-lens mode-locked laser resonator. The end mirrors are represented by
dashed lines.}%
\end{figure}

As an example, we choose here the spatiotemporal pulse propagation in a
Kerr-lens mode-locked laser resonator. The setup, shown in Fig.\thinspace1,
consists of a nonlinear Kerr medium and linear resonator arms to its left and
right, which contain an element with negative dispersion and a focusing
element.\cite{jir03} For the resonator arms, we assume a focal length
$f_{L}=5\,\mathrm{cm}$ and a group delay dispersion $S_{1}=S_{2}%
=-75\,\mathrm{fs}^{2}$. For the Kerr medium, the material parameters of
Ti:sapphire are used, with $\mathcal{D}=60\,\mathrm{fs}^{2}/\,\mathrm{mm}$,
$\delta=k_{0}n_{2I}=0.25\,\mathrm{\mu m}/\,\mathrm{MW}$, $n_{0}=1.76$,
$f_{0}=375\,\mathrm{THz}$ and $\mathcal{B}=1/\left(  2n_{0}k_{0}\right)
=36.2\,\mathrm{nm}$. The lengths in the resonator are given by $L_{1}%
=80\,\mathrm{cm}$, $L_{2}=110\,\mathrm{cm}$, $l_{1}=5.05\,\mathrm{cm}$,
$L=0.25\,\mathrm{cm}$ and $l_{2}=5.2\,\mathrm{cm}$.

\begin{figure}[th]
\centering\includegraphics{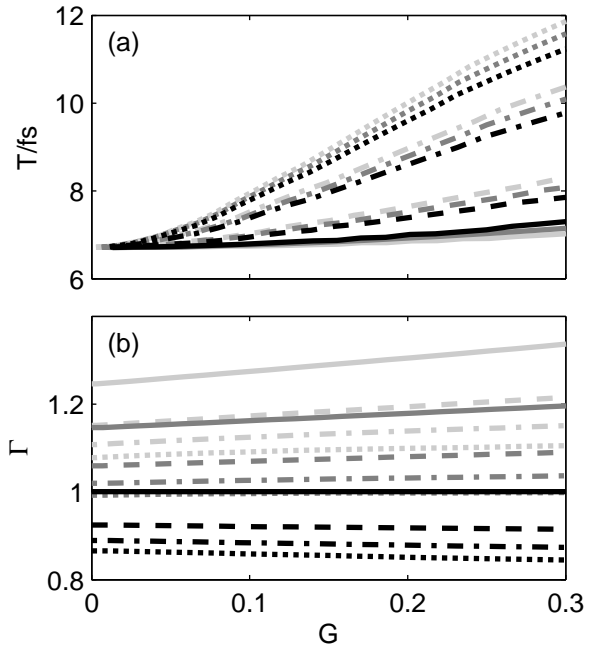}\caption{Steady-state pulse solution of
the setup shown in Fig. 1 for a fixed intracavity pulse energy of
$20\,\mathrm{nJ}$ at the right end mirror. Shown is a) the pulse duration $T$
at right end mirror and b) the stability factor $\Gamma$ as a function of the
gain per roundtrip $G$. Results for $\Delta_{\omega}/\left(  2\pi\right)
=40$, $45$, $56\,\mathrm{THz}$ and $\Delta_{\omega}\rightarrow\infty$ are
represented by dotted, dash-dotted, dashed and solid lines, respectively.
Transverse gain widths of $\Delta_{x}=10\,\mathrm{\mu m}$, $\Delta
_{x}=20\,\mathrm{\mu m}$ and $\Delta_{x}\rightarrow\infty$ are marked by light
gray, dark gray and black colors.}%
\end{figure}

In Ref. 4, the Gaussian solutions for the setup in Fig.\thinspace1 are
obtained taking into account only the energy-conserving effects, i.e.,
neglecting gain and loss. Here, a Gaussian gain profile in the Kerr medium is
added, characterized by the parameters $\hat{g}$, $\Delta_{\omega}$ and
$\Delta_{x}=\Delta_{y}$, see Eq.\thinspace(\ref{gg}). The output coupling is
taken into account by normalizing the intracavity pulse energy at the right
end mirror to $E=20\,\mathrm{nJ}$. In the following, we examine the dependence
of the solution on the gain parameters. Fig.\thinspace2 shows the Gaussian
pulse duration $T$ at the right end mirror and the stability factor $\Gamma$
as a function of the roundtrip gain $G=\Delta E/E$ for different values of
$\Delta_{x}$ and $\Delta_{\omega}$. $\Gamma$ is defined as the ratio between
the gain per roundtrip in pulsed and cw operation, $\Gamma=G/G_{\mathrm{cw}}$,
where $G_{\mathrm{cw}}$ is the roundtrip gain for the Gaussian steady-state
beam solution of the linear resonator. The stability factor serves as an
indicator for the suppression of the cw lasing in pulsed operation, a value of
$\Gamma>1$ indicating stable pulsed operation.\cite{jir03}

Fig.\thinspace2(a) shows the pulse duration $T$ at the right end mirror. The
pulse duration increases for strong spectral gain filtering, i.e., for a
strong roundtrip gain $G$ in combination with small gain bandwidth
$\Delta_{\omega}$. Also a small transverse gain width $\Delta_{x}$ leads for a
fixed value of $G$ to increased spectral filtering, mainly because of the
increased peak gain $\hat{g}$. Thus, minimum pulse durations can be obtained
in a setup with a small gain and weak outcoupling at the end mirrors.
Fig.\thinspace2(b) shows the stability parameter $\Gamma$. Due to the high
peak power, the pulsed solution experiences self-focusing and thus transverse
contraction in the Kerr medium, leading to an increased transverse overlap
with the gain. On the other hand, the cw solution is not affected by spectral
gain filtering. Thus, $\Gamma$ reaches maximum values for a small transverse
gain width $\Delta_{x}$ and a broad gain bandwidth $\Delta_{\omega}$.

The results for vanishing gain, $G\rightarrow0$, coincide with the ones
obtained in Ref. 4, taking into account only the energy-conserving effects.
For a small gain of a few percent, the energy-conserving dynamics is still a
good approximation, as can be seen from Fig.\thinspace2. However, the
equations of motion with gain have the distinct advantage that the system is
attracted by its stable solutions, while for the energy-conserving equations,
additional boundary conditions have to be introduced to find the steady-state solutions.\cite{jir03}

As pointed out in Ref. 4, higher-order dispersive effects, which are not
included here, can considerably affect the pulse shape. In addition, the
Gaussian approximation fails for excessive self-focusing and self-phase
modulation, as well as strong nonparabolic gain aperturing and filtering.
Under extreme conditions, the nonlinear Schr\"{o}dinger equation itself, as
given in Eq.\thinspace(\ref{nls}),\ loses its validity.

\section{\label{conclusion}CONCLUSION}

In conclusion, we have studied the spatial, temporal and spatiotemporal
optical propagation in Kerr media with transverse and spectral gain filtering
by applying the variational principle. Based on the Kantorovitch method, we
derived the Gaussian equations of motion for parabolic and general gain
profiles. By reformulating the variational equations as differential equations
for the $q$ parameters, we could extract $ABCD$ matrices for the Kerr effect
and a general transverse and spectral gain profile. As an example, we studied
the steady-state spatiotemporal Gaussian pulse dynamics in a Kerr-lens
mode-locked\ laser resonator.

The equations of motion can be solved efficiently with a standard differential
equation solver and allow for a quick simulation of the Gaussian optical
propagation through gain media with Kerr nonlinearity. By iterative solution
of these equations, the steady-state pulse or beam shape in a laser resonator
can be obtained. Further effects, like a parabolic refractive index profile as
generated by thermal lensing in a laser rod, can easily be considered by
additional $ABCD$ matrices. Gain saturation can be taken into account by
complementing the equations of motion with suitable gain saturation equations.

\section*{ACKNOWLEDGMENT}

This work was supported by ONR and DARPA under contract N00014-02-1-0717 and
HR0011-05-C-0155, respectively.

\appendix\renewcommand{\theequation}{A\arabic{equation}} \setcounter{equation}{0}

\section*{APPENDIX A: DERIVATION OF THE VARIATIONAL EQUATIONS}

In this appendix, we derive from Eq.\thinspace(\ref{nls}) the equations of
motion for the spatiotemporal Gaussian pulse parameters, using the
Kantorovitch method.\cite{cer98} The conservative Lagrangian is given by
\cite{jir02}%
\begin{align}
\mathcal{L} &  =\frac{\mathrm{i}}{2}\left(  U^{\ast}\frac{\partial U}{\partial
z}-U\frac{\partial U^{\ast}}{\partial z}\right)  +\mathcal{D}\left|
\frac{\partial U}{\partial t_{\mathrm{r}}}\right|  ^{2}-\mathcal{B}\left|
\frac{\partial U}{\partial x}\right|  ^{2}-\mathcal{B}\left|  \frac{\partial
U}{\partial y}\right|  ^{2}\nonumber\\
&  +\frac{\delta}{2}\left|  U\right|  ^{4},\label{L(u)_wt}%
\end{align}
while the non-conservative process is described by the expression $Q$, given
in Eq.\thinspace(\ref{Q}) and Eq.\thinspace(\ref{Q_g}), respectively. For the
envelope $U$, we insert the test function Eq.\thinspace(\ref{Uwt_wt}). The
Euler-Lagrange equations for the real parameter functions $f=A,\phi
,T,w_{x},w_{y},b,a_{x},a_{y}$ are then given by \cite{cer98}
\begin{equation}
\frac{\partial\left\langle \mathcal{L}\right\rangle }{\partial f}%
-\frac{\mathrm{d}}{\mathrm{d}z}\frac{\partial\left\langle \mathcal{L}%
\right\rangle }{\partial f^{\prime}}=R_{f},\label{Euler}%
\end{equation}
with the reduced Lagrangian
\begin{equation}
\left\langle \mathcal{L}\right\rangle =\int_{-\infty}^{\infty}\int_{-\infty
}^{\infty}\int_{-\infty}^{\infty}\mathcal{L}\mathrm{d}t_{\mathrm{r}}%
\mathrm{d}x\mathrm{d}y
\end{equation}
and the non-conservative term%
\begin{equation}
R_{f}=2\Re\left\{  \int_{-\infty}^{\infty}\int_{-\infty}^{\infty}\int
_{-\infty}^{\infty}Q\frac{\partial U^{\ast}}{\partial f}\mathrm{d}%
t_{\mathrm{r}}\mathrm{d}x\mathrm{d}y\right\}  .\label{Rf}%
\end{equation}
Using the definition of the Fourier transform in Eq.\thinspace(\ref{four}) and
Parseval's theorem, we can with $\mathcal{F}_{t}\left\{  Q\right\}
=\mathrm{i}g\tilde{U}$ express Eq.\thinspace(\ref{Rf}) as%
\begin{equation}
R_{f}=\frac{1}{\pi}\Re\left\{  \mathrm{i}\int_{-\infty}^{\infty}\int_{-\infty
}^{\infty}\int_{-\infty}^{\infty}g\tilde{U}\frac{\partial\tilde{U}^{\ast}%
}{\partial f}\mathrm{d}\omega\mathrm{d}x\mathrm{d}y\right\}  .\label{Rf2}%
\end{equation}
Since $g$ is real, we have $R_{A}=R_{w_{x}}=R_{w_{y}}=0$. Furthermore, for a
parabolic gain profile with the gain term Eq.\thinspace(\ref{Q}), we obtain%
\begin{equation}
R_{\phi}=E\left[  2g_{0}-g_{x}w_{x}^{2}-g_{y}w_{y}^{2}-g_{\omega}\Omega
^{2}\right]  ,\label{R_phi1}%
\end{equation}%

\begin{equation}
R_{T}=4g_{\omega}EbT^{-1},
\end{equation}%
\begin{equation}
R_{b}=\frac{1}{2}ET^{2}\left[  2g_{0}-g_{x}w_{x}^{2}-g_{y}w_{y}^{2}+g_{\omega
}\left(  T^{-2}-12T^{2}b^{2}\right)  \right]  ,
\end{equation}%
\begin{equation}
R_{a_{x}}=\frac{1}{2}Ew_{x}^{2}\left[  2g_{0}-3g_{x}w_{x}^{2}-g_{y}w_{y}%
^{2}-g_{\omega}\Omega^{2}\right]  ,\label{R_ax1}%
\end{equation}
and a corresponding expression for $R_{a_{y}}$, where the pulse energy is
given by%
\begin{equation}
E=\pi^{3/2}A^{2}Tw_{x}w_{y}.\label{E_wt}%
\end{equation}
From Eqs.\thinspace(\ref{R_phi1}) - (\ref{R_ax1}), we can extract equations
for $g_{\omega}$, $g_{x}$, $g_{y}$\ and $g_{0}$,%
\begin{equation}
g_{\omega}=R_{T}T/\left(  4Eb\right)  ,\label{gw1}%
\end{equation}%
\begin{equation}
g_{x}=\left(  R_{\phi}w_{x}^{2}/2-R_{a_{x}}\right)  /\left(  Ew_{x}%
^{4}\right)  ,
\end{equation}%
\begin{equation}
g_{y}=\left(  R_{\phi}w_{y}^{2}/2-R_{a_{y}}\right)  /\left(  Ew_{y}%
^{4}\right)  ,
\end{equation}%
\begin{equation}
g_{0}=\left(  R_{\phi}/E+g_{x}w_{x}^{2}+g_{y}w_{y}^{2}+g_{\omega}\Omega
^{2}\right)  /2.\label{g01}%
\end{equation}
This enables us to express any gain profile formally through effective
parabolic gain parameters, which depend on both the gain function and the
pulse parameters. Inserting Eq.\thinspace(\ref{Rf2}) into Eqs.\thinspace
(\ref{gw1}) -- (\ref{g01}), we arrive at Eq.\thinspace(\ref{gwt}).

Setting $f=A$ in Eq.\thinspace(\ref{Euler}) yields%
\begin{equation}%
\begin{array}
[c]{c}%
-2\phi^{\prime}-\left(  b^{\prime}T^{2}+a_{x}^{\prime}w_{x}^{2}+a_{y}^{\prime
}w_{y}^{2}\right)  +\mathcal{D}\left(  \frac{1}{T^{2}}+4b^{2}T^{2}\right)  \\
-\mathcal{B}\left(  \frac{1}{w_{x}^{2}}+4a_{x}^{2}w_{x}^{2}\right)
-\mathcal{B}\left(  \frac{1}{w_{y}^{2}}+4a_{y}^{2}w_{y}^{2}\right)
+\delta\frac{A^{2}}{\sqrt{2}}=0,
\end{array}
\label{EulerA_wt}%
\end{equation}
and for $f=\phi$, we get%
\begin{equation}
E^{\prime}=E\left[  2g_{0}-g_{x}w_{x}^{2}-g_{y}w_{y}^{2}-g_{\omega}\left(
\frac{1}{T^{2}}+4b^{2}T^{2}\right)  \right]  .\label{Eulerphi_wt}%
\end{equation}
The Euler-Lagrange equations for the pulse duration and the beam widths are
given by%
\begin{equation}%
\begin{array}
[c]{c}%
-\phi^{\prime}+\frac{1}{2}\Big[-b^{\prime}T^{2}-3a_{x}^{\prime}w_{x}^{2}%
-a_{y}^{\prime}w_{y}^{2}+\mathcal{D}\left(  \frac{1}{T^{2}}+4b^{2}%
T^{2}\right)  \\
-\mathcal{B}\left(  -\frac{1}{w_{x}^{2}}+12a_{x}^{2}w_{x}^{2}\right)
-\mathcal{B}\left(  \frac{1}{w_{y}^{2}}+4a_{y}^{2}w_{y}^{2}\right)  \Big
]+\frac{\delta}{4\sqrt{2}}A^{2}=0,
\end{array}
\label{Eulerwx_wt}%
\end{equation}%
\begin{equation}%
\begin{array}
[c]{c}%
-\phi^{\prime}+\frac{1}{2}\Big[-b^{\prime}T^{2}-a_{x}^{\prime}w_{x}^{2}%
-3a_{y}^{\prime}w_{y}^{2}+\mathcal{D}\left(  \frac{1}{T^{2}}+4b^{2}%
T^{2}\right)  \\
-\mathcal{B}\left(  \frac{1}{w_{x}^{2}}+4a_{x}^{2}w_{x}^{2}\right)
-\mathcal{B}\left(  -\frac{1}{w_{y}^{2}}+12a_{y}^{2}w_{y}^{2}\right)  \Big
]+\frac{\delta}{4\sqrt{2}}A^{2}=0,
\end{array}
\label{Eulerwy_wt}%
\end{equation}%
\begin{equation}%
\begin{array}
[c]{c}%
-\phi^{\prime}+\frac{1}{2}\Big[-3b^{\prime}T^{2}-a_{x}^{\prime}w_{x}^{2}%
-a_{y}^{\prime}w_{y}^{2}+\mathcal{D}\left(  -\frac{1}{T^{2}}+12b^{2}%
T^{2}\right)  \\
-\mathcal{B}\left(  \frac{1}{w_{x}^{2}}+4a_{x}^{2}w_{x}^{2}\right)
-\mathcal{B}\left(  \frac{1}{w_{y}^{2}}+4a_{y}^{2}w_{y}^{2}\right)  \Big
]+\frac{\delta}{4\sqrt{2}}A^{2}=4g_{\omega}b,
\end{array}
\label{EulerT_wt}%
\end{equation}
and for the chirp parameters, we obtain%
\begin{align}
&  \frac{1}{2}E^{\prime}+E\frac{w_{x}^{\prime}}{w_{x}}-4\mathcal{B}%
Ea_{x}\nonumber\\
&  =E\left[  g_{0}-\frac{3g_{x}}{2}w_{x}^{2}-\frac{g_{y}}{2}w_{y}%
^{2}-2g_{\omega}\left(  \frac{1}{4T^{2}}+b^{2}T^{2}\right)  \right]
,\label{Eulerax_wt}%
\end{align}%
\begin{align}
&  \frac{1}{2}E^{\prime}+E\frac{w_{y}^{\prime}}{w_{y}}-4\mathcal{B}%
Ea_{y}\nonumber\\
&  =E\left[  g_{0}-\frac{g_{x}}{2}w_{x}^{2}-\frac{3g_{y}}{2}w_{y}%
^{2}-2g_{\omega}\left(  \frac{1}{4T^{2}}+b^{2}T^{2}\right)  \right]
,\label{Euleray_wt}%
\end{align}%
\begin{align}
&  \frac{1}{2}E^{\prime}+E\frac{T^{\prime}}{T}+4\mathcal{D}Eb\nonumber\\
&  =E\left[  g_{0}-\frac{g_{x}}{2}w_{x}^{2}-\frac{g_{y}}{2}w_{y}%
^{2}+2g_{\omega}\left(  \frac{1}{4T^{2}}-3T^{2}b^{2}\right)  \right]
.\label{Eulerb_wt}%
\end{align}
Inserting Eq.\thinspace(\ref{Eulerphi_wt}) into Eqs.\thinspace
(\ref{Eulerax_wt}) -- (\ref{Eulerb_wt}) results in Eqs.\thinspace(\ref{w}) and
(\ref{T}). Multiplying Eqs.\thinspace(\ref{Eulerwx_wt}), (\ref{Eulerwy_wt})
and (\ref{EulerT_wt}) by $2$ and subtracting Eq.\thinspace(\ref{EulerA_wt})
from each equation yields Eqs.\thinspace(\ref{a}) and (\ref{b}).
Eq.\thinspace(\ref{A_wt}) can be obtained from Eq.\thinspace(\ref{Eulerphi_wt}%
) by inserting Eqs.\thinspace(\ref{E_wt}), (\ref{w}) and (\ref{T}).
Multiplying Eq.\thinspace(\ref{EulerA_wt}) by a factor of $2\frac{1}{2}$ and
subtracting Eqs.\thinspace(\ref{Eulerwx_wt}), (\ref{Eulerwy_wt}) and
(\ref{EulerT_wt}) yields Eq.\thinspace(\ref{phi_wt}).

\renewcommand{\theequation}{B\arabic{equation}} \setcounter{equation}{0}

\section*{APPENDIX B: DERIVATION OF THE EQUATIONS FOR THE q PARAMETERS}

The spatiotemporal dynamics is described by coupled equations of motion for
$q_{x}$, $q_{y}$, $q_{t}$ and $\hat{U}$. Differentiation of Eq.\thinspace
(\ref{qp}) with respect to $z$ yields with Eqs.\thinspace(\ref{w}) and
(\ref{a}) the differential equation for $q_{p}$,%
\begin{equation}
q_{p}^{\prime}=-\left(  \frac{2\mathrm{i}g_{p}}{k_{0}}-\frac{2c_{a}\delta
}{k_{0}w_{p}^{2}}\left|  \hat{U}\right|  ^{2}\right)  q_{p}^{2}+\frac{1}%
{n_{0}},\label{qp_wt}%
\end{equation}
with $p=x,y$ and $w_{p}=\left(  k_{0}\Im\left\{  q_{p}^{-1}\right\}  \right)
^{-1/2}$. The differential equation for $q_{t}$ is obtained by differentiating
Eq.\thinspace(\ref{qt}) with respect to $z$ and inserting Eqs.\thinspace
(\ref{T}) and (\ref{b}):%

\begin{equation}
q_{t}^{\prime}=-\frac{2c_{a}\delta}{\omega_{0}T^{2}}\left|  \hat{U}\right|
^{2}q_{t}^{2}+2\mathcal{D}\omega_{0}+2\mathrm{i}g_{\omega}\omega
_{0},\label{qt_wt}%
\end{equation}
with $T=\left(  -\omega_{0}\Im\left\{  q_{t}^{-1}\right\}  \right)  ^{-1/2}$.
Differentiating Eq.\thinspace(\ref{U}) with respect to $z$ and inserting
Eqs.\thinspace(\ref{A_wt}) and (\ref{phi_wt}) yields with Eqs.\thinspace
(\ref{qp}) and (\ref{qt}) the equation of motion for the complex amplitude,%
\begin{equation}
\frac{\hat{U}^{\prime}}{\hat{U}}=\frac{A^{\prime}}{A}+i\phi^{\prime}%
=-\frac{\mathcal{B}k_{0}}{q_{x}}-\frac{\mathcal{B}k_{0}}{q_{y}}-\frac
{\mathcal{D}\omega_{0}}{q_{t}}-\mathrm{i}\frac{\omega_{0}g_{\omega}}{q_{t}%
}+g_{0}+\mathrm{i}c_{\phi}\allowbreak\delta\left|  \hat{U}\right|
^{2}.\label{U_wt}%
\end{equation}
The purely spatial beam propagation, Eq.\thinspace(\ref{Uq_w}), is described
by Eqs.\thinspace(\ref{qp_wt}) and (\ref{U_wt}) with $\mathcal{D}=g_{\omega
}=0$, and the temporal pulse propagation, Eq.\thinspace(\ref{Uq_t}), is
described by Eqs.\thinspace(\ref{qt_wt}) and (\ref{U_wt}) with $\mathcal{B}%
=0$. The nonlinearity coefficients are given in Table 1.

In the $q$ parameter formalism, discrete optical elements are represented by
$2\times2$ matrices%
\begin{equation}
M_{s}=\left(
\begin{array}
[c]{cc}%
A_{s} & B_{s}\\
C_{s} & D_{s}%
\end{array}
\right)  .\label{Mp}%
\end{equation}
For the spatiotemporal pulse propagation, each optical element is
characterized by three matrices, i.e., $s=x,y,t$. The transformation law for
the propagation through an optical element extending from position $z_{1}$ to
$z_{2}$,%
\begin{equation}
q_{s}\left(  z_{2}\right)  =\frac{A_{s}q_{s}\left(  z_{1}\right)  +B_{s}%
}{C_{s}q_{s}\left(  z_{1}\right)  +D_{s}},\label{qp_trans}%
\end{equation}
is valid for both the spatial and temporal $q$ parameters.\cite{sie86,dij90}
The amplitude at position $z_{2}$ is given by%
\begin{align}
\hat{U}\left(  z_{2}\right)   &  =\tau\hat{U}\left(  z_{1}\right)  \left[
A_{x}+B_{x}/q_{x}\left(  z_{1}\right)  \right]  ^{-1/2}\nonumber\\
&  \times\left[  A_{y}+B_{y}/q_{y}\left(  z_{1}\right)  \right]
^{-1/2}\left[  A_{t}+B_{t}/q_{t}\left(  z_{1}\right)  \right]  ^{-1/2}%
\label{Uwt_trans}%
\end{align}
for a spatiotemporal Gaussian pulse,%
\begin{align}
\hat{U}\left(  z_{2}\right)   &  =\tau\hat{U}\left(  z_{1}\right)  \left[
A_{x}+B_{x}/q_{x}\left(  z_{1}\right)  \right]  ^{-1/2}\nonumber\\
&  \times\left[  A_{y}+B_{y}/q_{y}\left(  z_{1}\right)  \right]
^{-1/2}\label{Uw_trans}%
\end{align}
for a spatial beam, and%
\begin{equation}
\hat{U}\left(  z_{2}\right)  =\tau\hat{U}\left(  z_{1}\right)  \left[
A_{t}+B_{t}/q_{t}\left(  z_{1}\right)  \right]  ^{-1/2}\label{Ut_trans}%
\end{equation}
for a purely temporal pulse, with the on-axis transmission $\tau=\exp\left(
\int\alpha\mathrm{d}z\right)  $, where $\alpha$ is the complex on-axis
transmission coefficient. The propagation equations Eqs.\thinspace
(\ref{qp_wt})--(\ref{U_wt}) can be obtained by dividing the Kerr medium into
small sections of length $\delta z$, and representing each section by $ABCD$
matrices of the form Eq.\thinspace(\ref{Ms}). From Eqs.\thinspace
(\ref{qp_trans}) and (\ref{Uwt_trans}),\ we obtain%
\begin{equation}
q_{s}\left(  z+\delta z\right)  =\frac{q_{s}\left(  z\right)  +B_{s}^{\prime
}\delta z}{q_{s}C_{s}^{\prime}\delta z+1}\approx-q_{s}^{2}\left(  z\right)
C_{s}^{\prime}\delta z+q_{s}\left(  z\right)  +B_{s}^{\prime}\delta z
\end{equation}
and%
\begin{align}
\hat{U}\left(  z+\delta z\right)   &  =\hat{U}\left(  z\right)  \exp\left(
\alpha\delta z\right)  \left(  1+B_{x}^{\prime}\delta z/q_{x}\right)
^{-1/2}\nonumber\\
&  \times\left(  1+B_{y}^{\prime}\delta z/q_{y}\right)  ^{-1/2}\left(
1+B_{t}^{\prime}\delta z/q_{t}\right)  ^{-1/2}\nonumber\\
&  \approx\hat{U}\left(  z\right)  \left(  1-\frac{B_{x}^{\prime}\delta
z}{2q_{x}}-\frac{B_{y}^{\prime}\delta z}{2q_{y}}-\frac{B_{t}^{\prime}\delta
z}{2q_{t}}+\alpha\delta z\right)  .
\end{align}
In the limit $\delta z\rightarrow0$, this results in Eqs.\thinspace
(\ref{q2_w}) and (\ref{U2_wt}). Comparison with Eqs.\thinspace(\ref{qp_wt}),
(\ref{qt_wt}), and (\ref{U_wt}) yields the elements Eq.\thinspace(\ref{BC})
and the $\alpha$\ given in Eq.\thinspace(\ref{alpha}). The equations for the
purely spatial or temporal dynamics can be derived in an analogous manner.

\end{document}